\documentclass[preprint2]{aastex}

\usepackage{amsmath}
\usepackage{graphicx}
\usepackage{epsfig,psfrag,epic,eepic}
\usepackage{textcomp}
\usepackage{times}

\slugcomment{For submission to the Astrophysical Journal Letters}

\shorttitle{The AK Sco Circumbinary Disk}
\shortauthors{Janson et al.}

\begin{document}

\title{\Large Detection of Sharp Symmetric Features in the Circumbinary Disk Around AK Sco\altaffilmark{*}}

\author{Markus Janson\altaffilmark{1},
Christian Thalmann\altaffilmark{2},  
Anthony Boccaletti\altaffilmark{3}, 
Anne-Lise Maire\altaffilmark{4,5}, 
Alice Zurlo\altaffilmark{6,7}, 
Francesco Marzari\altaffilmark{8}, 
Michael R. Meyer\altaffilmark{2}, 
Joseph~C. Carson\altaffilmark{9,4}, 
Jean-Charles Augereau\altaffilmark{10,11}, 
Antonio Garufi\altaffilmark{2},
Thomas Henning\altaffilmark{4}, 
Silvano Desidera\altaffilmark{5}, 
Ruben Asensio-Torres\altaffilmark{1},
Adriana Pohl\altaffilmark{4}
}

\altaffiltext{*}{Based on observations collected at the European Southern Observatory, Chile, under observing program 095.C-0346(B).}
\altaffiltext{1}{Department of Astronomy, Stockholm University, AlbaNova University Center, 106 91 Stockholm, Sweden}
\altaffiltext{2}{Institute for Astronomy, ETH Zurich, Wolfgang-Pauli-Strasse 27, CH-8093 Zurich, Switzerland}
\altaffiltext{3}{LESIA, Observatoire de Paris--Meudon, CNRS, Universit{\'e} Pierre et Marie Curie, Universit{\'e} Paris Didierot, 5 Place Jules Janssen, F-92195 Meudon, France}
\altaffiltext{4}{Max Planck Institute for Astronomy, K\"onigstuhl 17, 69117 Heidelberg, Germany}
\altaffiltext{5}{INAF--Osservatorio Astromonico di Padova, Vicolo dell'Osservatorio 5, 35122 Padova, Italy}
\altaffiltext{6}{N\'ucleo de Astronom\'ia, Facultad de Ingenier\'ia, Universidad Diego Portales, Av. Ejercito 441, Santiago, Chile}
\altaffiltext{7}{Millennium Nucleus ``Protoplanetary Disk'', Departamento de Astronom\'ia, Universidad de Chile, Casilla 36-D, Santiago, Chile}
\altaffiltext{8}{Dipartimento di Fisica, University of Padova, Via Marzolo 8, 35131 Padova, Italy}
\altaffiltext{9}{Department of Physics and Astronomy, College of Charleston, 66 George Street, Charleston, SC 29424, USA}
\altaffiltext{10}{Universit{\'e} Grenoble Alpes, IPAG, 38000 Grenoble, France}
\altaffiltext{11}{CNRS, IPAG, 38000 Grenoble, France}

\begin{abstract}\noindent
The Search for Planets Orbiting Two Stars (SPOTS) survey aims to study the formation and distribution of planets in binary systems by detecting and characterizing circumbinary planets and their formation environments through direct imaging. With the SPHERE Extreme Adaptive Optics instrument, a good contrast can be achieved even at small ($<$300~mas) separations from bright stars, which enables studies of planets and disks in a separation range that was previously inaccessible. Here, we report the discovery of resolved scattered light emission from the circumbinary disk around the well-studied young double star AK~Sco, at projected separations in the $\sim$13--40~AU range. The sharp morphology of the imaged feature is surprising, given the smooth appearance of the disk in its spectral energy distribution. We show that the observed morphology can be represented either as a highly eccentric ring around AK Sco, or as two separate spiral arms in the disk, wound in opposite directions. The relative merits of these interpretations are discussed, as well as whether these features may have been caused by one or several circumbinary planets interacting with the disk.

\end{abstract}


\section{Introduction}
\label{s:introduction}

AK~Sco is a spectroscopic F5+F5 binary in the Upper Centaurus Lupus (UCL) association with an estimated age of $\sim$10--20~Myr \citep[e.g.][]{song2012,pecaut2012}. Its distance is relatively uncertain, with the original Hipparcos catalogue \citep{perryman1997} giving a value of 145$^{+38}_{-25}$~pc while the newer \citet{vanleeuwen2007} reduction gives a value of 102$^{+26}_{-17}$~pc. However, interferometric measurements have independently provided a distance estimate of 141$\pm$7~pc \citep{anthonioz2015}, hence we use this latter value here. The system is classified as a HAeBe disk system from infrared excess \citep[e.g.][]{jensen1997,menu2015}, and variability has been observed at a range of wavelengths from interactions between the disk and the central binary \citep{manset2005,gomezdecastro2013a,gomezdecastro2013b}. Recently, the inner disk was resolved with near-infrared interferometry \citep{anthonioz2015} and the full disk was imaged in thermal radiation at moderate resolution with ALMA \citep{czekala2015}. As shown in previous studies of accretion onto AK~Sco \citep[e.g.][]{alencar2003}, the ALMA imaging confirms that the disk hosts a rather large quantity of gas (estimated mass of $7 \times 10^{-3}$~$M_{\rm sun}$), which is unusual given the progressed age of UCL.

The central binary has a semi-major axis of approximately 0.16~AU (1.11~mas at 141~pc) and the disk appears to have a gap with an inner rim at 0.58~AU (4.1~mas) from modelling of the interferometric visibilities \citep{anthonioz2015}. However, outside of this inner range, there is no clear evidence of additional gaps in the spectral energy distribution (SED) of the disk \citep{jensen1997}, hence one might naively expect the disk to appear continuous at larger separations in resolved imaging. Here, we report on the detection of sharp features in near-infrared imaging of the disk at projected separations of $\sim$13--40~AU, which contrast with this expectation. 

\section{Observations and Data Reduction}
\label{s:observations}

The observations presented here were acquired as part of the ongoing SPOTS program \citep{thalmann2014b}, which is primarily dedicated to direct imaging detection of circumbinary planets. Executed on 14 Apr 2015, these observations made use of ESO's newly commissioned SPHERE instrument \citep{beuzit2008} in the IRDIFS setting, in which the IRDIS dual-band imager and IFS integral field spectrograph are used simultaneously. IFS was set to YJ mode, covering wavelengths from 0.96~$\mu$m to 1.33~$\mu$m in steps of $\sim$0.01~$\mu$m, and for IRDIS the H2H3 mode was used, providing simultaneous dual-band imaging in two adjacent intermediate-width bands within the H-band range, centered on 1.593~$\mu$m and 1.667~$\mu$m, respectively. The N{\_}ALC{\_}YJH{\_}S coronagraph was used, setting an inner working angle of $\sim$92.5~mas. We employed pupil tracking during the observations, allowing for efficient angular differential imaging (ADI). A field rotation of 29$^{\rm o}$ was acquired between the first and last exposure of the observation. Detector dithering was used in order to optimize bad pixel removal for IRDIS. Every IRDIS frame was read out in four sub-integrations, each with a direct integration time of 16 seconds, and every IFS frame was read out in a single integration of 64 seconds. The total integration time in each instrument was 25.6 minutes.

All data were reduced using a combination of the Data Reduction and Handling package \citep[DRH, see][]{pavlov2008} and custom IDL routines \citep[e.g.][]{mesa2015}. ADI processing was applied to the data using both classical ADI \citep[cADI;][]{marois2006}, conservative LOCI \citep{thalmann2010}, and KLIP \citep{soummer2012} procedures. The centering of each frame was initially performed using the four satellite spots from a calibration sequence taken with a periodic modulation imposed on the deformable mirror adjacent to the observational sequence. All of the resulting images showed good consistency apart from the fact that the IRDIS images seemed shifted with respect to the IFS images by approximately 2 pixels. On close inspection of the data, it was found that while the satellite spots provided a center that was consistent with the location of the residual diffraction spot from the central star behind the coronagraphic mask in the IFS data, this was not the case for the IRDIS data. As a result, we deemed the satellite spots to be unreliable for the IRDIS data, and used the central spot directly for the centering instead. This provided a much better consistency between all data sets. 

\section{Results and Discussion}
\label{s:results}

\begin{figure*}[p]
\centering
\includegraphics[width=\textwidth]{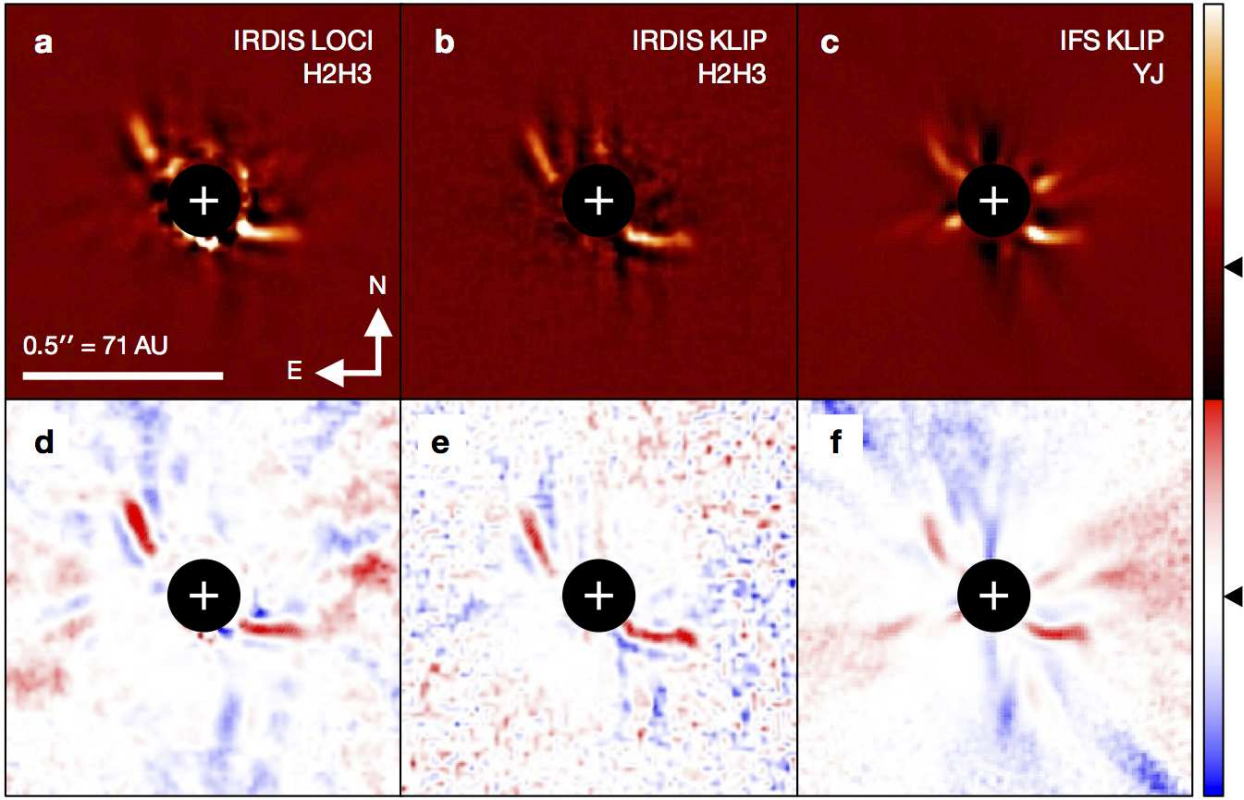}
\caption{SPHERE high-contrast images of AK Sco. \textbf{(a)} IRDIS data reduced with conservative LOCI, shown at a linear scale. \textbf{(b)} The same data reduced with KLIP. \textbf{(c)} IFS data reduced with KLIP and collapsed across the spectral dimension. \textbf{(d--f)} S/N maps corresponding to the three images. The color scale spans $[-4\sigma, +4\sigma]$. In both color scale bars, the black triangle marks the zero level.  All images show the two arms of the disk discussed in the text.}
\label{f:images}
\end{figure*}

The IFS and IRDIS images are shown in Fig.~\ref{f:images}. Two `arms' can be seen extending from each side of the central star in an approximately symmetrical fashion. These arms show up with every type of reduction in both the IRDIS and IFS data sets. We also verified that in contrast to the residual speckle pattern around the star, these arms retain a consistent morphology across all wavelength bands, as would be expected for features originating from the stellar Point Spread Function (PSF). Hence, we can conclude that the features are `real', i.e., they constitute scattered radiation from off-axis material in the circumbinary disk of AK~Sco. The entire disk structure visible in the images fits within a 0.1--0.3\arcsec\ radius around the star. Considering that instruments based on conventional AO systems have generally effectively been limited to studying separations only beyond $\sim$0.3\arcsec\ \citep[e.g.][]{janson2013}, this detection illustrates the considerable benefit of using Extreme AO for studies of the circumstellar environment.

In the following subsections, we will discuss two lines of interpretation for the observed structures: (1) an eccentric ring of material surrounding a gap, and (2) a pair of spiral arms propagating through an otherwise predominantly smooth disk.

\subsection{Eccentric ring interpretation}
\label{s:ring}

Structures similar to those revealed around AK~Sco are often found in disks that contain rings of material with gaps inside them \citep[e.g.][]{fitzgerald2007,buenzli2010,thalmann2014a}. Such gapped disks often manifest as bright crescents in high-contrast imaging due to anisotropic forward scattering enhancing the brightness of their near-side edge. Unlike these cases, AK~Sco does not have strong indications in its SED for any gap in the relatively wide separation range that we probe with these images \citep{jensen1997}. Nonetheless, given the qualitative similarities in morphology and the many degeneracies implicit to SED analysis, it makes sense to test if this line of interpretation can reproduce the data, and if so, what parameters the corresponding disk would have.

For the purpose of this analysis, we use the IFS data with a cADI reduction. While this provides marginally lower S/N than more sophisticated algorithms such as LOCI or KLIP for these data, it allows for a much more robust and rapid modelling of the ADI self-subtraction effects imposed on the observational data. We use the GRATER code \citep{augereau1999} to generate ring models with a range of parameters. 

Each model is subjected to the same cADI procedure as the data and then subtracted from the data. The $\chi^2$ in the region where the disk resides is then calculated in order to assess the quality of the fit. The model parameters that were varied initially were the peak radius of the dust belt $r$, the Henyey-Greenstein scattering index $g$, the inner and outer power-law slopes $\alpha_{\rm in}$ and $\alpha_{\rm out}$, the eccentricity $e$, the argument of periapsis $\omega$, and the inclination $i$. Rather than including the position angle as a parameter, it was determined individually on the basis of symmetry in the target image. The image was rotated in steps of 0.1$^{\rm o}$, and the left half of the image was flipped and subtracted from the right half with the ambition of finding the rotation angle that minimizes the residuals. We determined a position angle of 53.4$^{\rm o}$ with this procedure. 

Once a best fit was determined, we fixed the values of $\alpha_{\rm in}$ and $\alpha_{\rm out}$ since they had little impact on the fit quality, and evaluated covariances between each other pair of parameters. Because the residuals are strongly influenced by remnant speckle noise, it is not practical to define a stringent $\chi^2$ cut-off criterion, but instead we define the range of parameters that give an acceptable fit as those that fulfil the condition $\chi^2 < 2 \chi_{\rm min}^2$. This corresponds to the range of models that are visually acceptable. 

\begin{figure*}[p]
\centering
\includegraphics[width=\textwidth]{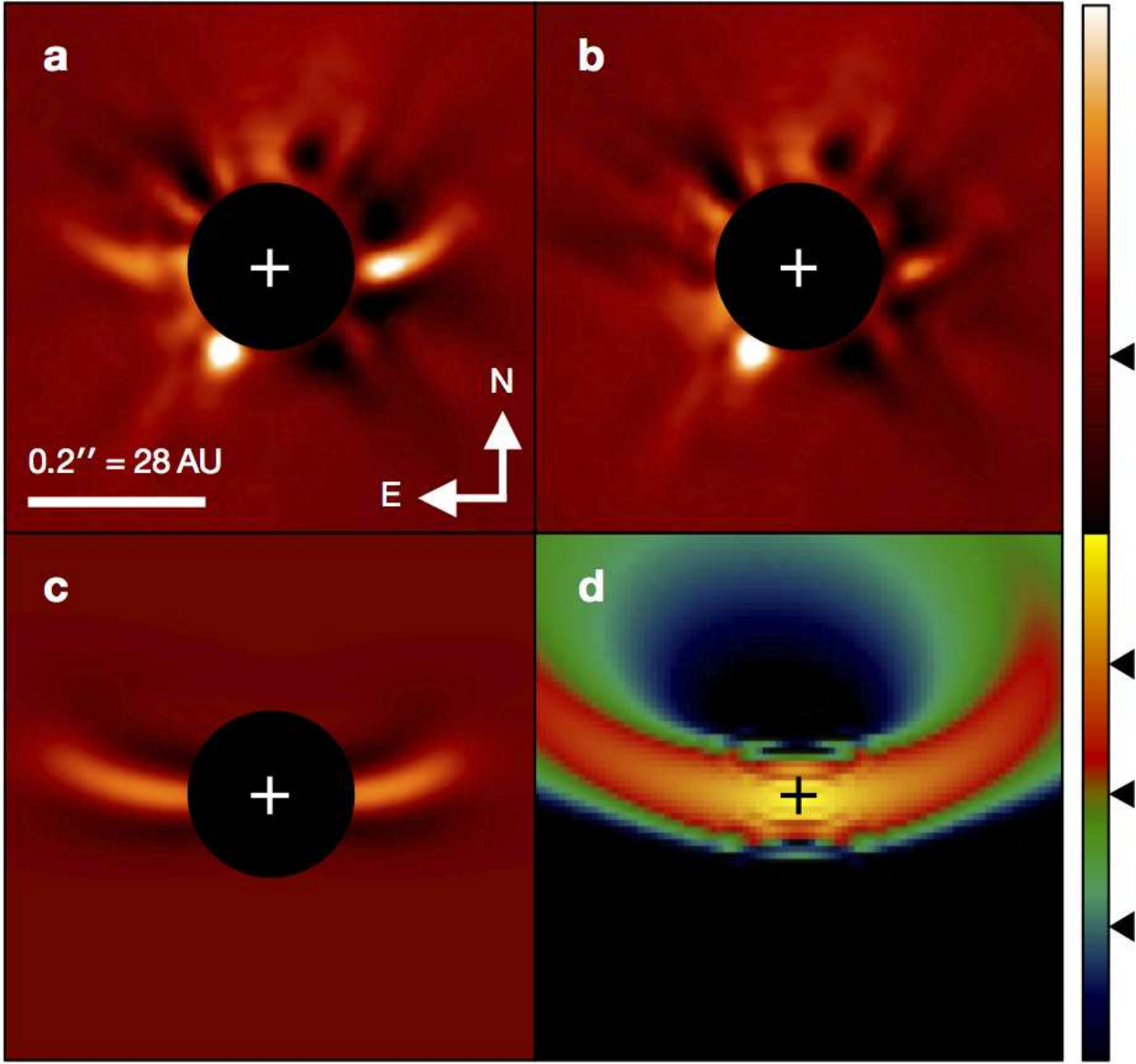}
\caption{Result of the model fitting of an eccentric ring with the GRATER code. The location of the central binary is marked with a plus sign in each panel. \textbf{(a)} The cADI-reduced IFS data without model subtraction. \textbf{(b)} The cADI-reduced data with the best-fit model subtracted. \textbf{(c)} The cADI-reduced best-fit model by itself. \textbf{(d)} The best-fit model with a logarithmic stretch, shifted to show the morphology of the full model. It should however be recalled that the apoapsis of the ring is very loosely constrained and could be much closer to or farther away from the star. The black triangles mark the zero level on the linear scale bar (top) and orders of magnitude on the logarithmic scale bar (bottom). }
\label{f:ring}
\end{figure*}

The best-fit model can be seen in Fig. \ref{f:ring}. It reproduces most of the observed features, except for the brightness asymmetry between the left and right arms, which cannot be reproduced by any model parameters. All acceptable fits have a high eccentricity. This is to be expected, since the location of the arms implies a strongly offset ellipse relative to the location of the central star, which is characteristic of a high eccentricity. The periapsis of close to 90$^{\rm o}$ (82.3--93.0$^{\rm o}$) is quite robust, as the morphology rapidly becomes significantly more asymmetric than the observed flux otherwise. It could be argued that this $\sim$90$^{\rm o}$ alignment constitutes a fine-tuning problem for the gap interpretation.

The $g$ index is virtually unconstrained by this procedure, with no positive values of $g$ being excluded in the fitting. It is also not strongly covariant with any other parameter. This is to be expected in these images, where a small region of the disk is probed and the feature shoots out from the star at a steep angle, such that a very small range of scattering angles are represented in the observed flux. The $i$ parameter is only weakly covariant with $r$ and $e$ and $\omega$ is not noticeably covariant with any of the other quantities. The only very strong covariance is between $r$ and $e$. The reason for this is that we only probe the ring close to its periapsis with no information about where the apoapsis resides, hence a higher $e$ can be compensated for by a larger $r$ as long as the periapsis distance $r_{\rm p} = r(1-e)$ remains similar. Indeed, while the semimajor axis of the ring can only be constrained within a factor of $\sim$10 as shown in table \ref{t:disk}, $r_{\rm p}$ can be constrained at least within a factor 3--4, with values of 3.5--12.0~AU providing acceptable residuals (the best fit is at 6.3~AU).

\begin{table}[p]
\caption{Disk parameters if interpreted as an eccentric ring, from GRATER fitting.}
\label{t:disk}
\centering
\begin{tabular}{lcc}
\hline
\hline
Quantity & Best fit & Range \\
\hline
$g$	&	0.15	&	0--1 	\\
$r$	&	126~AU	&	32--350~AU	\\
$i$	&	98.0$^{\rm o}$	&	93.9--103.1$^{\rm o}$	\\
$e$	&	0.95	&	0.63--0.99	\\
$\omega$	&	88.5$^{\rm o}$	&	82.3--93.0$^{\rm o}$	\\
$\alpha_{\rm in}$	&	7	&	Fixed	\\
$\alpha_{\rm out}$	&	$-$6	&	Fixed	\\
\hline
\end{tabular}
\end{table}

It appears that an eccentric gap can provide a reasonable match to the data, which would most likely imply the presence of a highly eccentric planet forcing the ring into an apsidally locked state. However, several arguments can be made against an eccentric gap as the explanation for the observed structure. Firstly, there is no sign of any gap in the continuum nor the gas lines in spatially resolved ALMA data of AK~Sco \citep{czekala2015}. The Full Width at Half Maximum (FWHM) of the ALMA imaging is $\sim$0.8\arcsec\ (113~AU at 141~pc), so the smaller gaps among acceptable fits to the SPHERE data may remain undetected in ALMA imaging. However, the velocity pattern of the gas shows no indication of any eccentric behaviour, and in fact an upper limit of the mean eccentricity of the gas disk is set at $e_{\rm mean} < 0.004$ in \citet{czekala2015}. The difference between this value and the $e > 0.66$ derived for the hypothesized eccentric gap appears rather large. There is also no sign of any gap at the separation range we are probing in the SED of AK~Sco. However, an eccentric gap edge entails material at a large range of separations and thus a large range of temperatures, and therefore will not result in any equally clear signature in an SED as a circular gap edge with a well-defined temperature cut-off. Finally, the derived inclination range of 94--103$^{\rm o}$ from the fitting is lower than the ALMA-derived inclination of 109.4$\pm$0.5$^{\rm o}$. We note that the latter is closer to the inclination of the binary orbit of 115$\pm$3$^{\rm o}$ \citep{anthonioz2015}. This may also imply a weakness in the eccentric ring hypothesis, although a warped disk could yield systematically different inclinations in different parts of the disk.

\subsection{Spiral arm interpretation}
\label{s:spirals}

Spiral arms have already been observed in a number of disks \citep[e.g.][]{muto2012,grady2013,garufi2013,boccaletti2013,benisty2015}. Such features can be induced through gravitational instability or through the influence of a planet or binary companion \citep[e.g.][]{dong2015a,dong2015b,pohl2015}. The features that we observe around AK Sco, if interpreted in a spiral arm context, would constitute two spiral arms that are wound in opposite directions (one unwinding clockwise and the other counter-clockwise). In fact, the central binary itself is predicted to impose spiral features in the disk of AK Sco \citep{gomezdecastro2013b}, but this is on a smaller scale than probed in our observation, and the two spiral arms simulated in \citet{gomezdecastro2013b} both unwind in the same direction, unlike our observed features.

In order to test whether the observed features can be explained as spiral arms, we fit spirals following the procedure of \citet{boccaletti2013}, based on the formalism of \citet{muto2012} which assumes a planet launching the spiral, located at position [$r_{\rm c}$,$\theta_0$] in polar coordinates. The IRDIS image was used for this purpose (both the KLIP and cADI reductions were used, and yielded mutually consistent results). This image was de-projected using an inclination angle of 70$^{\rm o}$ \citep{czekala2015}, and Gaussian profiles were fit at each position angle where an arm was visible, in order to construct a trace of the spine of each arm. We then fit the \citet{muto2012} relation to these traces. This relation takes the parameters $\theta_{\rm 0}$ and $r_{\rm c}$ as well as $h_{\rm c}$ (the disk aspect ratio at $r_{\rm c}$), and $\alpha$ and $\beta$ which are power-law indices for the radial dependencies of the angular frequency and sound speed in the disk, respectively. The $\alpha$ index was set to 1.5 and for $\beta$ we sequentially tried the values 0.1, 0.25, and 0.4, following the reasoning in \citet{boccaletti2013}. The two arms are individually fit with a separate spiral.

\begin{figure*}[p]
\centering
\includegraphics[width=12cm]{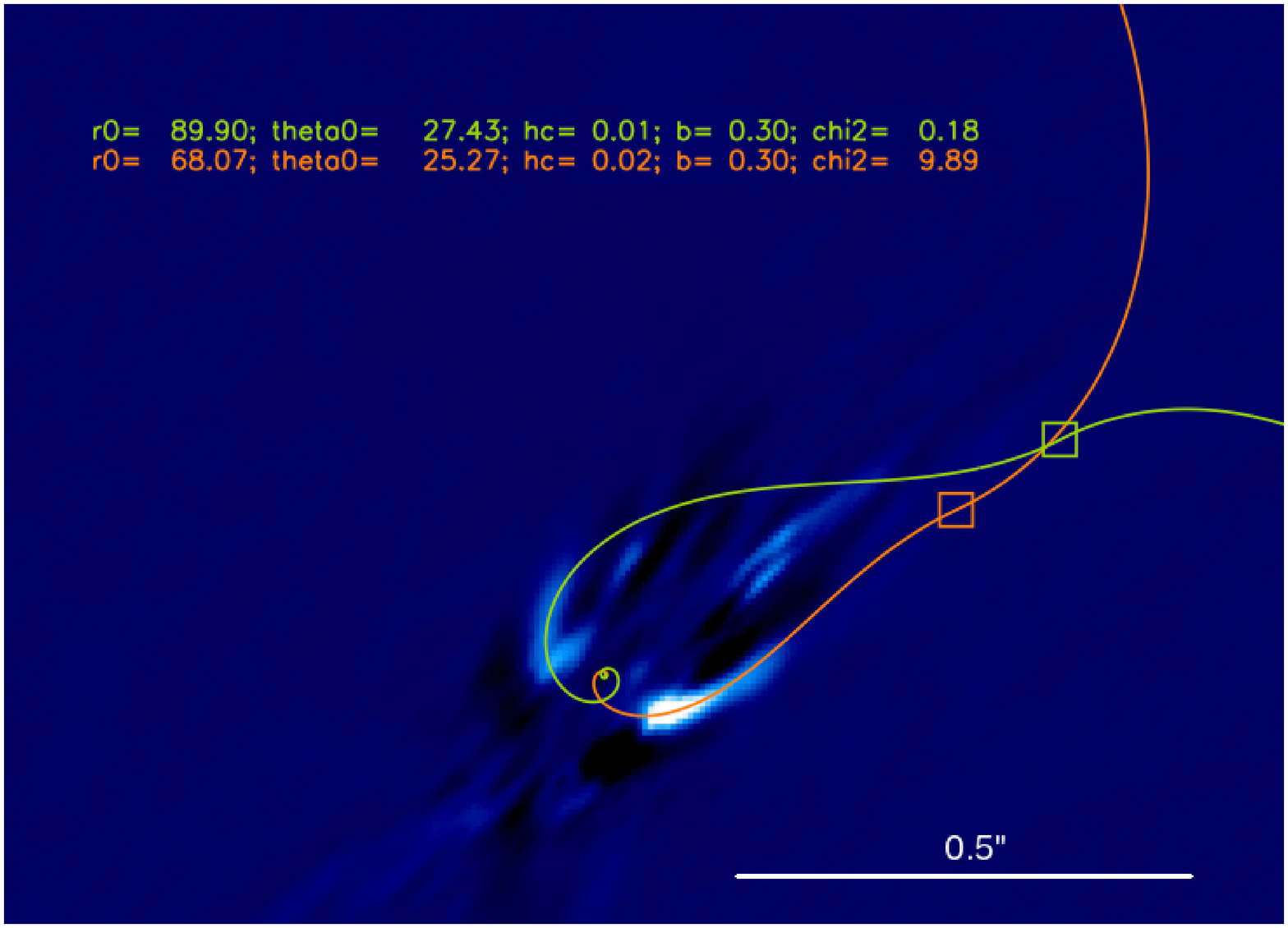}
\caption{Spiral fit to the de-projected image of AK~Sco. The lines are spiral fits to the observed arms, and the boxes mark the locations of any planets that would have launched such spirals. Spirals can be fit to the data, but with too large degeneracies to meaningfully constrain the parameters for the hypothesized planets.}
\label{f:spirals}
\end{figure*}

An example of a good fit is shown in Fig. \ref{f:spirals}. Two conclusions can readily be drawn from this procedure: The fact that there are well-fitting solutions indicates that spirals could provide an explanation for the observed morphology; and the fact that a vast range of degenerate parameters can yield acceptable fits indicates that few firm predictions can be made regarding the location of any planets that may be responsible for launching the spirals. The degeneracy is partly caused by the fact that only short sections of the putative spiral arms are detected in the image. Uncertainty in the inclination used for de-projection of the image also is also important in this regard. Despite these degenaracies, there are some general trends among fits yielding a small $\chi^2$: The two perturbing objects are always located toward the Northwest, typically at separations of 0.5--2\arcsec. All fits are consistent with a $h_{\rm c}$ of 0.01--0.04. In most cases, the two spiral tracks intersect, possibly leaving room for an interpretation of two spirals launched by a single perturber. Solutions in which the perturbers are interior to the arms do not yield satisfactory fits, as the resulting spirals are too tightly wound in such cases to reproduce the observed morphology.

We can test whether gravitational instability could conceivably account for spiral arms in the disk by estimating its $Q$ parameter. For this estimation we use the mass estimates for the gaseous disk component of $7 \times 10^{-3}$~$M_{\rm sun}$ and for the central binary as 2.49~$M_{\rm sun}$ from ALMA data \citep{czekala2015}. We adopt an outer radius of 160~AU based on the fact that the same ALMA data trace the disk out to approximately this separation, and assume a continuous density profile declining as $r^{-3/2}$. Using an isothermal approximation with a scale height of $h=0.037$, we then find that $Q > 30$ everywhere in the disk. While there are uncertainties in the precise disk parameters, this test implies that the AK~Sco disk is probably not gravitationally unstable. Rather, if the spiral arm hypothesis is valid for explaining the observed features, then it is more probable that the spirals would be launched by unseen planets in the disk. One peculiar aspect of the spiral arm scenario, as mentioned previously, is the fact that the two spiral arms are wound in opposite directions. This could perhaps imply that one of the hypothesized spiral-launching planets orbits the central binary in a retrograde fashion. Another aspect that could be seen as speaking against the spiral arm scenario is the fact that the features are so apparently symmetric, which there is no reason to assume should be the case for two independent spiral arms, but which would be a natural consequence for an gap edge.

\section{Summary}
\label{s:summary}

We have presented the discovery of spatially resolved scattered light from the circumbinary disk of AK~Sco. The morphology is unexpected given the smoothness of the SED of AK~Sco in the literature, and may imply the existence of a highly eccentric gap, or a set of two spiral arms, unwinding in opposite directions. Since the separation of the observed feature is a factor $\sim$100 larger than the binary semi-major axis, the binary itself probably doesn't directly affect the observed morphology. However, either of these scenarios may be indicative of circumbinary planetary companions in the disk. Planets are often inferred as probable causes of eccentricity in disks, since such a state is otherwise hard to attain \citep[e.g.][]{quillen2006}. Likewise, while gravitational instability appears unlikely as a direct cause of spirals in the low-mass disk of AK~Sco, planets launching the spirals remain a plausible scenario in the spiral arm interpretation. 

The features seen in the AK~Sco disk are reminiscent of those seen around HD~100546 \citep{boccaletti2013}, and this morphological likeness is strongly emphasized in newer SPHERE data of HD~100546 (A. Garufi et al., in prep.). These systematic similarities could imply that in contrast to the interpretations discussed above, another form of feature is observed, related either to the (partly ADI-affected) morphology or the scattering properties of moderately inclined disks. For instance, a flared inclined disk can produce similar morphologies \citep{watson2007}, although AK~Sco is classified as a class II HAeBe system \citep{menu2015}, which are generally characterised by a non-flared geometry. It should also be considered that ISM interactions with the disk can cause curved morphologies, as in cases such as HD~61005 and HD~32297 \citep{hines2007,debes2009}. New features such as those observed here are a natural outcome of the increased discovery range for disk studies that is opened by the new generation adaptive optics systems \citep[see e.g.][]{boccaletti2015}.

\acknowledgements
MJ gratefully acknowledges funding from the Knut and Alice Wallenberg foundation. CT and MRM acknowledge financial support from the SNSF via NCCR PlanetS and from the European Commission Marie Curie IEF grant 329875. ALM and SD acknowledge support via ``Progetti Premiali'' from MIUR. JC acknowledges support from the NSF under award 1009203. A.Z. acknowledges support from the Chilean Ministry of Economy, through grant ``Nucleus RC130007''.

\clearpage

\end{document}